# Numerical modelling of diffusion decomposition processes

N.M. Kiryukhin, A.V. Sagalovich, V.V. Sagalovich, V. N. Chabanovsky

A set of programs for the numerical simulation of the diffusion decomposition processes was developed by using simulation methods, kinetic and particle method. The complex has been validated on the model system Ni-Al by the growth of -phase separations. The results on the evolution of the distribution function and other characteristics of the ensemble, which in the zero volume fraction approximation are asymptotically in good agreement with the theory and the experiment, have been obtained. The peculiarity of the created program complex is the possibility of its adaptation to the description of the decomposition of multicomponent multiphase systems.
Fig. 6, list of references - 19 titles.

## Introduction

Diffusion decomposition of supersaturated solid solutions, which consists in the nucleation and subsequent growth of new phase precipitates, is the process, which leads to formation of heterogeneous structures with definite volume distribution of macro defects in a solid. This phenomenon is widely used in the technology of aging alloys for the production of high-strength steels, composite materials, heat-resistant alloys based on nickel, aluminum, etc.

At the same time, decomposition can present a serious danger in deterioration of material properties due to undesirable growth of various defects (precipitations, pores, etc.) during operation, actually limiting the resource life of the material. This, in fact, determines the enormous role that diffusion decomposition processes play in the formation and evolution of various structural-sensitive properties of solids. Without taking into account the distribution, interaction and kinetics of macrodefects it is impossible to give a correct physical picture of the behaviour of real materials under external influences and, consequently, it is impossible to program scientifically the creation of materials with improved parameters and to optimise their performance characteristics.

A consistent theory of diffusion decay at a late stage due to diffusion interaction of macrodefects is constructed in [1,2]. Even approximatedly formulated systems of equations are very complicated and can be fully analytically solved only asymptotically in time, not to mention more rigorous and more complex descriptions. This is especially true for multicomponent disperse systems, which admit an analytical description only in a small number of limiting cases.

In this connection, many physically interesting and practically important questions remain outside the limits of analytical research in the theory of diffusion decomposition - the study of kinetics of transition of a disperse system to an asymptotic state, construction of state and nonequilibrium diagrams of decay of multicomponent systems, investigation of cyclic thermomechanical treatment of disperse structures, etc.

Some of these problems can be successfully solved by numerical simulation methods. Despite their great promise, only a small number of works on diffusion decomposition have been performed with their help so far [3, 7].

The purpose of this work is to develop a set of programs for computers that allows one to numerically simulate the processes of diffusion decomposition of supersaturated solid solutions. The second section formulates the basic equations of the diffusion decomposition theory necessary for modeling and specifies two possible ways of their numerical solution. The third section describes the calculation algorithms and the capabilities of the software package. Finally, the fourth section presents the results of simulation of the decomposition process in the Ni-Al system and the software validation.

The peculiarity of the created complex is the possibility to adapt it to the description of the decomposition of multicomponent multiphase systems, which is the next stage of the research cycle.

## 2. Basic equations

Let us consider an ensemble of spherical emanations with radii $R_i$, dispersed in a matrix $M$. The equilibrium concentration $c_{R_i}$ at the surface of the radius $R_i$ extrusions depends substantially on $R_i$ [1]

$$c_{R_i} = c_\infty + \frac{\alpha}{R_i}, \qquad (1)$$

which leads to the flux $J_{R_i}$ from small emission into the matrix and from the matrix to large emission

$$J_{R_i} = D \frac{\partial c}{\partial r}\bigg|_{r=R_i}, \qquad (2)$$

where $\alpha = \dfrac{2\sigma V c_\infty}{kT}$, $c_\infty$ is the concentration of the saturated solution. $\sigma$ - interfacial surface energy, $V$ - volume of the solute atom, $D$ - its diffusion coefficient in the matrix.

To describe the behavior of a polydisperse ensemble, it is necessary to solve the complex diffusion problem

$$\frac{\partial c}{\partial r} = D \Delta c ; \quad c|_{r=R_i} = c_{R_i} \qquad (3)$$

When solving it one usually passes to description of diffusive growth of one selection with average concentration $\overline{c} = c|_{r=\infty}$ away from the selection, determined by the law of conservation of dissolved matter. The procedure of averaging is considered in detail in [8]. The change of sizes is taken into account by solving the continuity equation in the space of sizes of extractions. Following [1,2], we write in zero approximation the system of equations

$$\frac{dR}{dt} = V(R) = \frac{D}{R}(\overline{c} - c_R) = \frac{D}{R}(\Delta - \frac{\alpha}{R}) = \frac{D}{R}(\frac{1}{R_k} - \frac{1}{R}) \qquad (4а)$$

$$\frac{\partial f(R,t)}{\partial t} = \frac{\partial}{\partial R}[V(R)f(R,t)] = 0 \qquad (4б)$$

$$\overline{c} + \frac{4\pi}{3}\int_0^\infty f(R,t) R^3 dR = Q_0 \qquad (4в)$$

Here $f(R,t)$ is the function of emission size distribution at time $t$, normalized to the number of emission per unit volume

$$N(t) = \int_0^\infty f(R,t)dR, \qquad (4г)$$

$Q_0$ - total number of atoms of dissolved matter in the system (in the matrix and in the excretions); $\Delta = \overline{c} - c_\infty$ - supersaturation; $R_k = \dfrac{\alpha}{\Delta}$ -- critical radius of excretions.

Supplementing (4) with the initial conditions

$$f(R,t)|_{t=0} = f_0(R), \qquad (5а)$$

$$\overline{c}|_{t=0} = c_0 \qquad (5б)$$

we obtain a closed system of equations. Describing in the accepted approximations the late stage of diffusion decomposition.

The growth of macrodefects at this stage, when supersaturation is already very small, is essentially determined by the balance of point defects, the number of which in the solid solution becomes insufficient to provide simultaneous independent growth of all elements of the ensemble. A "diffusion interaction" arises between the macrodefects, when each macrodefect "feels" the self-consistent diffusion field of point defects determined by the whole ensemble. The developing competition for a power source leads to the fact that the growth of macrodefects occurs due to diffusive mass transfer of matter from smaller-sized macrodefects to larger ones.

Such an unusual behavior of an ensemble, as shown in the theory [1,2], causes the existence of a stable asymptotic state characterized by a universal macrodefects size distribution function and a cubic growth kinetics of average sizes $\left(R^3 \approx t\right)$. These conclusions turn out to be true also when generalizing the decay theory to much more complex multicomponent disperse systems [9,10]. Numerous experiments [11-14] confirm the correctness of the theoretical notions.

Note that the system of equations (3) corresponds to an approximation of the zero volume fraction of the discharges ($Q_0 \to 0$, $\overline{R} \langle\langle \overline{l}$, where $\overline{l}$ -the average distance between the discharges). The form of equations (4a) and (4b) changes at $Q_0 \neq 0$. In the first approximation, the changes $\frac{dR}{dt}$ at $Q_0 \neq 0$ are obtained in [8]. The algorithm for solving the modified equation (4d), taking into account the particle fusion mechanism, is given in (15), and numerical calculations are given in [16]. Using equations (4), it is possible to construct computational procedures for modeling polydisperse ensemble in the accepted approximations.

Different approaches to the construction of computational models are possible. One of them is based on the solution of the kinetic equation in partial derivatives (4b). The advantage of the method is the possibility to work with the integral characteristics of the ensemble and a high counting speed. However, in this model, it is not yet possible to give a consistent procedure for the following approximations due to the enormous mathematical difficulties arising in solving the more complex kinetic equation.

Another method is the particle method based on the solution of the diffusion problem (3) with appropriate boundary conditions. The particle method, allowing a more detailed description of an ensemble, in principle allows one to eliminate many difficulties related to approximations accepted in the theory, but, unfortunately, technical capabilities of modern computers are insufficient for an exact description of behavior of an ensemble even of several particles, which forces one to introduce again various approximations.

Despite these limitations, both methods can be useful for solving problems of the diffusion decay theory.

The software package developed by us uses both the kinetic method and the particle method, which significantly expands the modeling possibilities.

**3. Algorithms of calculations.**

Due to sufficient smoothness of the distribution function $f(R,t)$ in the entire domain of determination, the continuity equation (4b) can be solved using finite-difference schemes. We have used the Lax-Wendroff method [17], which ensures good convergence and stability if the grid is chosen correctly.

Let the distribution function at a point in time be defined in the domain $0 \langle R \langle R_k$. Let us cover this region with a grid with nodes and consider the values of the distribution function in the nodes of this grid.

Let the distribution function at time $t_1$ be defined in the region k. Let us cover the given region with a grid with nodes and consider the values of the distribution function in the nodes of this grid.

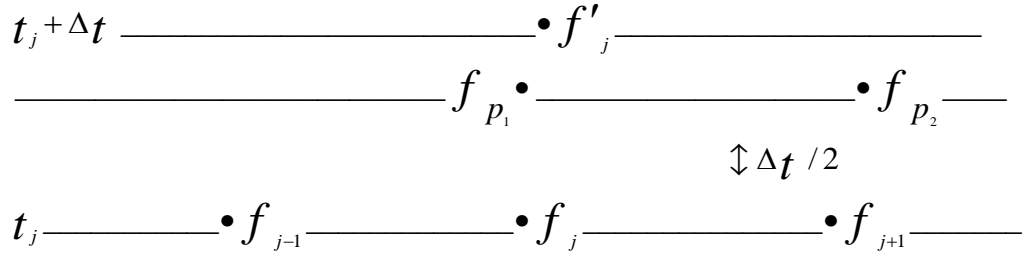

Fig. 1. The scheme of the Lax-Wendroff method

The Lax-Wendroff method is a two-step method. To calculate the values $f'_j$ ($0 \langle j \langle n$), the values of the distribution function at the three grid points of the previous time layer are used $f_{j-1}$, $f_j$ and $f_{j+1}$. First, we calculate the values of the distribution function at two intermediate points $p_1 = (R_{j-1} + R_j)/2$ and $p_2 = (R_j + R_{j+1})/2$ at the point in time $t_j + \Delta t/2$, then using the calculated values at the intermediate points we directly calculate $f'_j$.

We use the following difference schemes to replace equation (4b):
(a) on the intermediate layer

$$\frac{f_{p_1} - \frac{f_{j-1} + f_j}{2}}{\Delta t / 2} + \frac{V(R_j) f_j - V(R_{j-1}) f_{j-1}}{h} = 0$$

$$\frac{f_{p_2} - \frac{f_{j+1} + f_j}{2}}{\Delta t / 2} + \frac{V(R_{j+1}) f_{j+1} - V(R_j) f_j}{h} = 0$$

where

$$f_{p_1} = \frac{f_{j-1} + f_j}{2} + \frac{\Delta t}{2h} \left[ V(R_j) f_j - V(R_{j-1}) f_{j-1} \right] \qquad (6a)$$

$$f_{p_2} = \frac{f_{j+1} + f_j}{2} + \frac{\Delta t}{2h}\left[V(R_{j+1})f_{j+1} - V(R_j)f_j\right] \qquad (6b)$$

b) on the layer $t_1 + \Delta t$

$$\frac{f'_j - f_j}{\Delta t} + \frac{V(R_j + h/2)f_{p_2} - V(R_j - h/2)f_{p_2}}{h} = 0$$

whence

$$f'_j = f_n - \frac{f_n V(R_n) - f_{n-1} V(R_{n-1})}{h}\Delta t \qquad (7)$$

for all $0 \langle j \langle n$.

It is proved in [17] that this method has accuracy of order $h^2$ and is stable when a time step satisfying the condition

$$\Delta t / 2h \le 1/[V_{max}].$$

The Lax-Wendroff method makes it possible to calculate the values of the distribution function, based on equation (7), on the next time layer for all grid points in the domain of determination $f$, except for the boundary one. If on the left boundary at $r = 0$, $f_0 = 0$, the change $h\ V(R_k)\ f(R_k)$ should be specified separately. Let us use equation (4a) and take into account that at small ones $h\ V(R_k)$ it practically does not change. In this case the values $f'_n(t_1 + \Delta t)$ will be calculated by formula

$$f'_n = f_n - \frac{f_n V(R_n) - f_{n-1} V(R_{n-1})}{h}\Delta t \qquad (8)$$

In addition to calculating the distribution function value at the rightmost point of the grid, you should also take into account that the right boundary of the distribution function definition changes with every time step. Therefore the program continuously monitors the change in the distance of the rightmost point of the distribution function definition from the rightmost point of the grid for each time interval $\Delta t$. If, after the next time step, this distance exceeds the value of the grid step $h$, then the number of grid nodes, i.e., the area of the function definition and thus , used in equations (6) - (8), changes. The choice of the time step depends (inversely) on the maximum rate of change in the radius of separation. The growth rate reaches its maximum values at small radii. Therefore, the following procedure is implemented to speed up the counting.

1. The area where the distribution function is defined is divided into two subareas. Partitioning criteria can be different, for example, a given number of grid points in the first subregion or a given percentage of substance in it. Suppose in this case that the boundary point between subareas is a point $m$ (i.e., the first region is defined at $R_j$, $j = 1.....m$, , , and the second region is defined at $R_j$, $j = m...n$).

2. Calculate $V_{max}$ and $\Delta t_2$ for the second region and correspondingly new values of $f'_j$ $j = m,...n - 1$.

3. The multiplicity coefficient $M = \dfrac{\Delta t_2}{\Delta t_1}$ is calculated $V_{max}$ and $\Delta t_1$ for the first region as well, and an iterative process of calculating $f'_j$, $j = 1,...m$ by equations (6)-(8) is performed $M$ once.

The use of such an algorithm allows speeding up the modeling process by a factor of 8 to 20.

### 3.2. Particle method.

Choose a large enough volume element $\Delta W$ in the matrix $M$ and divide it by a cubic lattice into $N_i$ cells of size $L \langle\langle \overline{R}$. In the center of each cell we place a graduation with radius $R_i = R_i(\vec{r},t)$ ($\vec{r}_i$ -radius-vector of the center of $i$-one selection in the chosen coordinate system, Using Monte Carlo method, the selections are thrown into the cells from the given initial distribution $f_0$, so that normalization (4g) is performed. To eliminate the ordering effect, the coordinates of the centers are randomized, but so that the selection does not go beyond the cell and the distance between the selections in neighboring cells $\bar{l}$ significantly exceeds their radii ($(\bar{l}\rangle\rangle \overline{R}$). This choice makes it possible to further adapt the program to the solution of the diffusion problem in the non-zero volume fraction approximation (the approximation of the selections $\overline{R} \sim \bar{l}$).

Since the kinetic method operates with equations (4) obtained in the self-consistent diffusion field approximation, the same approximation (4a, 4c) is used for the particle method to compare the simulation results of both methods.

By discretization of equations (4a), (4c) we have to calculate the ensemble state $\{R_i^{n+1}\}$ at (N+1) step

$$R_i^{n+1} = R_i^n + \int_{t_n}^{t_{n+1}} V(R_i) dt,$$

where $V(R_i)$ in the accepted approximations is given by equation (4a) or refined by a more rigorous solution of the diffusion problem (3).

Using the first-order Euler method for approximation of integrals, which consists in replacing the integrand function on the segment by the value of this function at the moment, determining the supersaturation from the equation of balance of matter (4c), we can fully determine the state of the ensemble at the (n+1) step. For the numerical stability of this difference scheme, the time step at each time layer must satisfy the condition 2(Rn)2/D. The accuracy of such a scheme is of the order of .

Applying an implicit Euler method the essence of which consists in the replacement of the integrand function by the half-sum of its values at (n} and (n+1) steps we can increase the accuracy of calculation Rin+1 up to (t )2 .

Moreover, the stability of this scheme is absolute and does not depend on the choice of step. Some difficulties in this method are related to the fact that Rin+1 at (n+1} steps should be determined from an implicit equation, which leads to an increase in counting time.

An even higher accuracy (of the order of ( t)4) can be achieved using the Runge-Kut method, in which the function value is approximated by a four-point polynomial of the fourth power. The scheme also has high numerical stability. Its disadvantages are the same as those of the implicit scheme.

The program provides for the use of all three computational schemes. For sufficiently large selections, we select the maximum time step close to the stability boundary and use the first-order Euler scheme. For small separations, the accuracy of the separations and stability are improved by switching to an implicit scheme. Finally, calculations for the smallest separations are performed by the Runge-Kutta method.

At any time step, the evolution of histograms, central moments (mean size, dispersion, asymmetry, kurtosis), as well as total number of particles, oversaturation, and volume fraction of extractions can be visualized if necessary, which are compared with the corresponding theoretical asymptotic values using matstatistics methods.

### 4. Results

The initial data corresponding to the $Ni-Al$ system in which the $-\gamma'$ phase $-Ni_3Al$ evolution grows during annealing were chosen for modeling. This system is convenient as an object of modeling, first, because the $\gamma'$-phase emission in $Ni$ can be represented as one-component emission, which allows us to compare the results of modeling with the theory [1,2], and second, this system has been studied in detail experimentally [18,19].

A normal distribution and a rectangular step are chosen as the initial distribution function for testing the program.

The initial data used for the simulation are as follows:

$C_\infty = 12{,}0747 \cdot 10^{-2}$

$C_0 = 12{,}039 \cdot 10^{-2}$

$D = 1{,}81 \cdot 10^{-12}$ см²/с

$\sigma = 12{,}0747 \cdot 10^{-2}$ эрг/см².

The initial emission number is $10^5$ in the simulated volume, which corresponds to the emission density in the experiments [18] with which the comparison was made.

Figure 2-6 shows the simulation results of both kinetic and particle methods, which are in good agreement with each other. Over time, as can be seen in Fig. 2, the original distribution function transforms, approaching more and more asymptotically in time to the universal distribution function ($P(u)$ ($u = $ R/Rk) predicted by theory [1,2]. Not only qualitative but also quantitative agreement with the theory is found in the asymptotic behavior of the other ensemble characteristics ($R^3 \sim t$, $\Delta \sim t^{-1/3}$, $N \sim t^{-1}$) (Figs. 3-6).

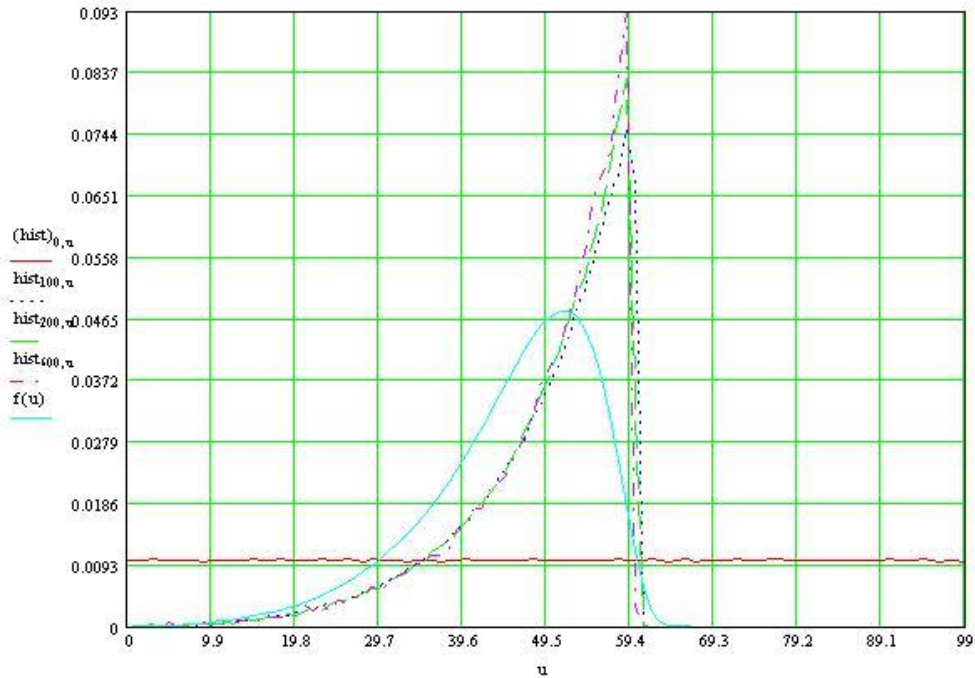

Fig. 3. The transformation of the distribution function in comparison with the universal distribution function (R/Rk), predicted by the theory.

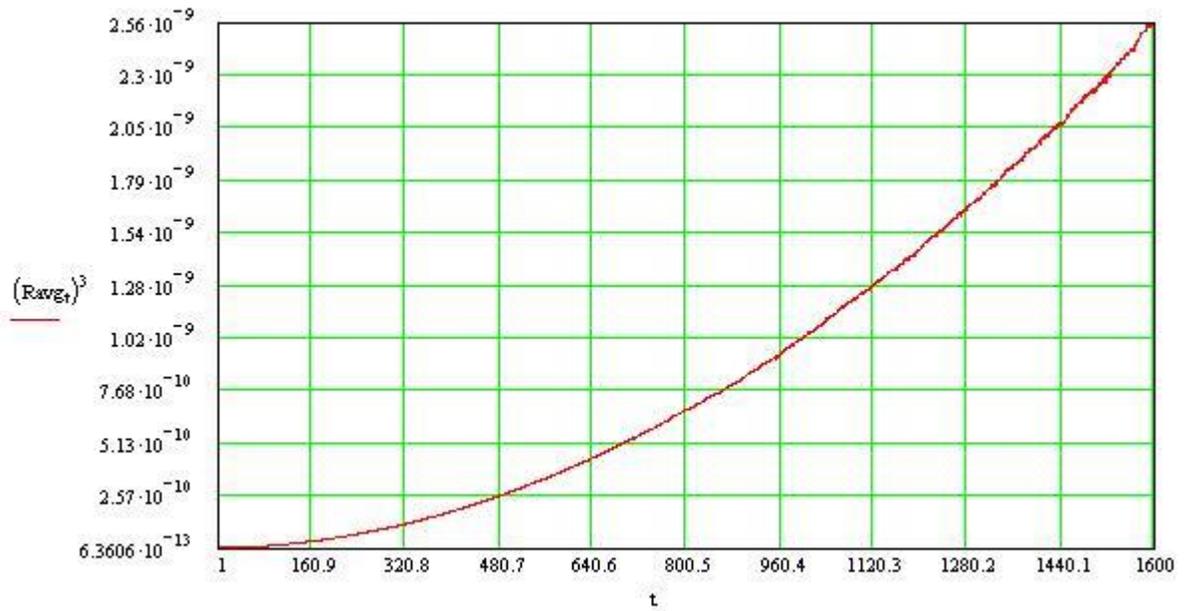

Fig. 4. Kinetics $(R_{avg})^3 = f(t)$.

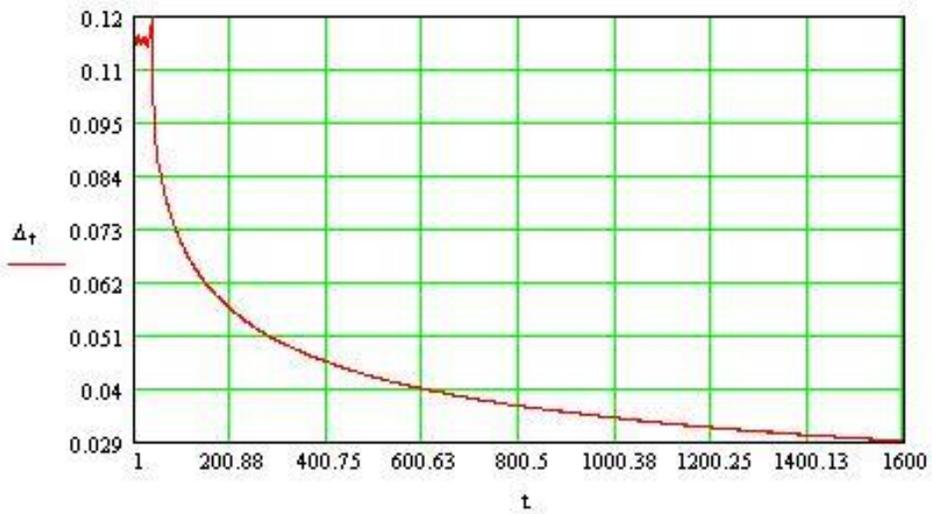

Fig. 5. Evolution of saturation $\Delta \sim t^{-1/3}$.

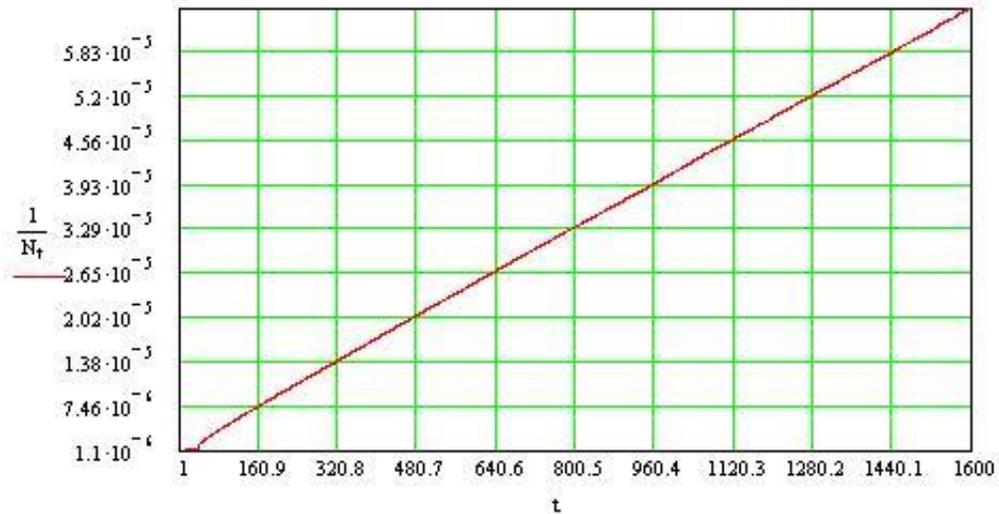

Fig. 6. Залежність $N^{-1}(t)$.

These results are in good agreement with experimental data [18,19].

**5. Conclusions**

1. We have developed a set of programs for the numerical simulation of the processes of diffusion decomposition of disperse systems, including two methods of simulation: the kinetic method and the particle method.

2. Numerical experiments on the growth of the precipitates in a model system $Ni - Al$ have been carried out.

3. The results on the evolution of the distribution function and other characteristics of the ensemble are obtained which asymptotically agree with the theory and the experiment in the zero volume fraction approximation.

4. Diffusion decay modeling by the kinetic and particle method yields well consistent results.

**List of references**